# A Comparative Study of SIP Overload Control Algorithms

Yang Hong, Changcheng Huang, James Yan

*Dept. of Systems and Computer Engineering, Carleton University, Ottawa, Canada*
*E-mail: {yanghong, huang}@sce.carleton.ca, jim.yan@sympatico.ca*

**ABSTRACT**

Recent collapses of SIP servers in the carrier networks indicates two potential problems of SIP: (1) the current SIP design does not easily scale up to large network sizes, and (2) the built-in SIP overload control mechanism cannot handle overload conditions effectively. In order to help carriers prevent widespread SIP network failure effectively, this chapter presents a systematic investigation of current state-of-the-art overload control algorithms. To achieve this goal, this chapter first reviews two basic mechanisms of SIP, and summarizes numerous experiment results reported in the literatures which demonstrate the impact of overload on SIP networks. After surveying the approaches for modeling the dynamic behaviour of SIP networks experiencing overload, the chapter presents a comparison and assessment of different types of SIP overload control solutions. Finally it outlines some research opportunities for managing SIP overload control.

## 1. INTRODUCTION

Internet telephony is experiencing rapidly growing deployment due to its lower-cost telecommunications solutions for both consumer and business services. Session Initiation Protocol (SIP) (Rosenberg et al., 2002) has become the main signaling protocol to manage multimedia sessions for numerous Internet telephony applications such as Voice-over-Internet Protocol (IP), instant messaging and video conferencing. 3rd Generation Partnership Project (3GPP) has adopted SIP as the basis of its IP Multimedia Subsystem (IMS) architecture (3GPP WG, 2011). With the 3G (3rd Generation) wireless technology being adopted by more and more carriers, most cellular phones and other mobile devices are starting to use or are in the process of supporting SIP for multimedia session establishment (Faccin, Lalwaney, & Patil, 2004).

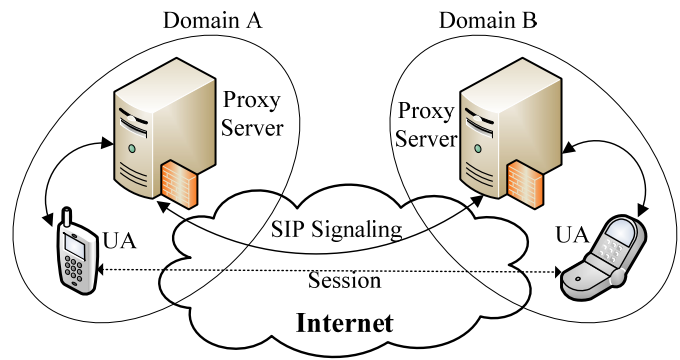

*Figure 1. Simplified architecture of a SIP network*

Figure 1 illustrates a simplified architecture of a SIP network. A SIP network consists of two types of basic elements: User Agent (UA) and Proxy Server (P-Server) (Rosenberg et al., 2002). A user agent can act as a user agent client (UAC) or as a user agent server (UAS). A P-server not only acts as the contact point with UA for core network service access but also provides routing for the signaling messages. SIP is responsible for establishing, modifying and terminating sessions for multimedia communication among multiple UAs.

RFC 5390 (Rosenberg, 2008) identifies the various reasons that may cause server overload in a SIP network. These include but are not limited to poor capacity planning, dependency failures, component failures, avalanche restart, flash crowds, denial of service attacks, etc. In general, anything that may trigger a demand burst or a server slowdown can cause server overload and lead to overload propagation and server crash, thus bringing down the whole SIP network.

The objective of this chapter is to present a systematic investigation of current state-of-the-art SIP overload control algorithms which aim at preventing server crashes in carrier networks. In order to provide a better knowledge of the major cause of SIP network collapse, the next section reviews two basic mechanisms of SIP, and describes the existing works on the performance study of the SIP overload. The third section surveys the related SIP modeling and analysis which can help network planners, operators, and researchers to understand how server overloading and widespread SIP network failure may happen under short-term demand bursts or server slowdowns. The forth section makes a comparative study of different types of SIP overload control solutions, thus helping carriers choose the appropriate solutions to avoid potential SIP network collapse (e.g., Skype outage (Ando, 2010) or VoIP outages in British Telecom, Vonage and Wanadoo (Materna, 2006)) in different overload situations. Finally, some future works for the SIP overload control are discussed.

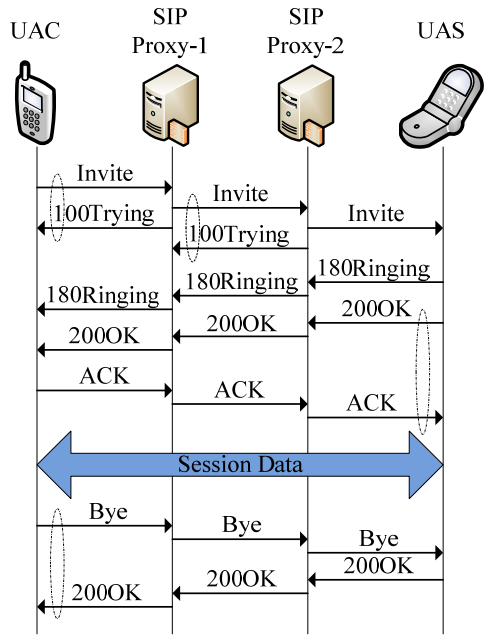

*Figure 2. A typical procedure of session establishment*

## 2. SIP OVERVIEW

SIP works in the application-layer for session establishment. Figure 2 depicts a typical procedure of a session establishment. To set up a call, a UAC sends an "Invite" request to a UAS via the two proxy servers. The proxy server or the UAS returns a provisional "100Trying" response to confirm the receipt of the "Invite" request. The UAS returns an "180Ring" response after confirming that the parameters are appropriate. It also evicts a "200OK" message to answer the call. The UAC sends an "ACK" response to the UAS after receiving the "200OK" message. Finally the call session is established and the media communication is created between the UAC and the UAS through the SIP session. The "Bye" request is generated to finish the session, thus terminating the communication.

## 2.1. SIP Retransmission Mechanism

SIP works independently of the underlying transport layer where Transmission Control Protocol (TCP) and User Datagram Protocol (UDP) may be deployed. SIP introduces a retransmission mechanism to maintain its reliability (Govind, Sundaragopalan, Binu, & Saha, 2003; Rosenberg et al., 2002). In practice, a SIP sender uses timeout to detect message losses. One or more retransmissions would be triggered if the corresponding reply message is not received in predetermined time intervals.

SIP RFC 3261 (Rosenberg et al., 2002) suggests that the SIP retransmission mechanism should be disabled for hop-by-hop transaction when running SIP over TCP to avoid redundant retransmissions at both SIP and TCP layer (Rosenberg et al., 2002). However, recent experimental evaluation on SIP-over-TCP overload behaviour by Shen & Schulzrinne (2010) demonstrates that TCP flow control mechanism cannot prevent SIP overload collapse for time-critical session-based applications due to lack of application context awareness at the transport layer. Other experiments (e.g., Hilt & Widjaja, 2008; Noel & Johnson, 2007) also indicate that the throughput with SIP-over-TCP exhibits similar overload collapse behaviour as that with SIP-over-UDP. Nahum, Tracey, & Wright (2007) claim that using TCP to deliver SIP messages degrades server performance from 43% (under stateful proxy with authentication) to 65% (under stateless proxy without authentication) when compared with using UDP.

Nearly all vendors choose to run SIP over UDP instead of TCP for the following reasons (Hilt & Widjaja, 2008; Hong, Yang, & Huang, 2004; Noel & Johnson, 2007; Shen & Schulzrinne, 2010; Stevens, 1994): (1) The overhead of state management such as three-way handshake prevents TCP from real time application which is a critical requirement for SIP protocol; (2) SIP works at application layer while TCP works at transport layer. Even if TCP can provide reliability at transport layer, SIP messages can still be dropped or corrupted at the application layer; (3) Designed for preventing congestion caused by bandwidth exhaustion, the complex TCP congestion control mechanism provides little help for SIP overload which is caused by CPU constraint.

In order to provide the reliable recovery of the message loss, SIP has two types of message retransmission: (a) a sender starts the first retransmission of the original message at $T_1$ seconds, the time interval doubling after every retransmission (exponential backoff), if the corresponding reply message is not received. The last retransmission is sent out at the maximum time interval $64 \mathrm{x} T_1$ seconds. Thus there is a maximum of 6 retransmissions. Default value of $T_1$ is 0.5s. The hop-by-hop "Invite"-"Trying" transaction shown in Figure 2 follows this rule (Rosenberg et al., 2002); (b) a sender starts the first retransmission of the original message at $T_1$ seconds, the time interval doubling after every retransmission but capping off at $T_2$ seconds, if the corresponding reply message is not received. The last retransmission is sent out at the maximum time interval $64 \mathrm{x} T_1$ seconds. Thus there is a maximum of 10 retransmissions. Default value of $T_2$ is 4s. The end-to-end "OK"-"ACK" and "Bye"-"OK" transactions shown in Figure 2 follows this rule (Rosenberg et al., 2002).

## 2.2. SIP Built-in Overload Control Mechanism

When the message arrival rate exceeds the message processing capacity at a SIP server, overload occurs and the queue increases, which may result in a long queuing delay and trigger unnecessary message retransmissions from its upstream servers. Such redundant retransmissions increase the CPU loads of both the overloaded server and its upstream servers. In this way, overload may propagate from server to server in a network, and eventually bring down the entire network (Hilt & Widjaja, 2008).

The SIP protocol provides a basic overload control mechanism through a 503 (Service Unavailable) response (Rosenberg et al., 2002). When a temporary overload occurs at a SIP downstream receiving server, it can decline to forward any request by sending a 503 response to its corresponding upstream sending server. The overloaded downstream server can insert a Retry-After header into the 503 response message, which specifies the duration during which the upstream server should not send any further requests to amplify the overload. After the duration expires, the upstream server can attempt to forward the requests to detect whether the overload is cancelled at the downstream server or not. Without a Retry-After header, a 503 response only rejects the current request, while all other requests can still be forwarded to the overloaded downstream server. This will amplify the overload by making the overloaded server continue spending CPU resources to reply with further 503s. After receiving a 503 response, the upstream server can try to re-route the request to other alternate server.

## 2.3. Experimental Results of the Impact of SIP Overload

In IP-telephony converged networks, customers expect timely responses to their service requests. Therefore, all the signaling elements must meet the requirements of a session setup, which means that SIP servers should process messages within real-time constraints. The main processing requirements of SIP elements were discussed and a number of performance metrics were described by Cortes, Ensor, & Esteban (2004). Performance measurement based on four different implementations demonstrates that parsing, string handling, memory allocation, and thread architecture have a significant impact on SIP elements (Cortes, Ensor, & Esteban, 2004). Noel & Johnson (2007) have investigated the impact of the overload on a reference SIP-based VoIP network through numerous simulations. Under the two cases of no controls and applying 503 Retry overload control mechanism, the comparison of the message goodput and call blocking probability demonstrates that the built-in 503 retry control can increase the message goodput from 32% to 56% (Noel & Johnson, 2007).

Without additional overload control mechanisms applied, the comprehensive study of SIP overload performance by Hilt & Widjaja (2008) demonstrates the two facts: (1) the servers using the current SIP protocol are vulnerable to overload and overload collapse, when SIP servers are deployed in different domains on a large scale (e.g., different types of session control functions in IMS); (2) a significant drop in goodput can be observed, if the server capacity is reached, when SIP runs over TCP and hop-by-hop retransmission is disabled. In addition, Linux experimental result provided by Nahum, Tracey, & Wright (2007) shows that when overload occurs at a SIP server, the response time increases sharply rather than linearly in proportion to the load, thus deteriorating the performance very quickly.

All these experimental results indicate the potential collapse of a SIP network due to the overload, which has already happened in the real carrier networks (e.g., VoIP outages in British Telecom, Vonage and Wanadoo, see Materna, 2006).

## 3. MODELING AND ANALYSIS OF SIP OVERLOAD

Modeling has become an efficient approach to study the properties of a signalling protocol. In order to find the root cause of SIP overload collapse, different analytical models and fluid models have been proposed to analyze the statistical characteristic or dynamic behaviour of SIP.

### 3.1. Analytical Models For Stable Signaling Systems

Prior to SIP becoming the dominant signaling protocol in the Internet, modeling and analysis of general signaling protocols have been done in the past. Ji, Ge, Kurose, & Towsley (2007) compared soft-state and hard-state signaling protocols based on the probabilities of inconsistent states among different servers along a signaling path. The states of a signaling protocol were modelled as a Markov chain under the assumptions: (a) there is no sudden signaling demand surge; (b) both arrival rate and service rate are Poisson process; (c) the signaling network is stable (Ji, Ge, Kurose, & Towsley, 2007). Instead of providing a model for a specific signaling protocol such as SIP, Ji, Ge, Kurose, & Towsley (2007) described five different signaling approaches that incorporate various hard-state and soft-state mechanisms, and qualitatively discussed the factors that influence system performance. All signaling protocols can be classified into five categories: pure soft-state signaling (SS), soft-state signaling with explicit removal (SS+ER), soft-state signaling with reliable trigger messages (SS+RT), soft-state signaling with reliable trigger/removal message (SS+RTR, e.g., SIP), and hard-state signaling (HS).

Based on the same assumptions made by Ji, Ge, Kurose, & Towsley (2007), some models have been created for signaling protocols (e.g., Lui, Misra, & Rubenstein, 2004; May, Bolot, Jean-Marie, & Diot, 1999; Raman & McCanne, 1999; Zaim et al., 2003). Raman & McCanne (1999) presented an analytical model for soft-state protocol and defined an associated consistency metric to evaluate the tradeoffs and performance of soft state communication. Queuing analysis and simulation were performed to study the data consistency and performance tradeoffs under a range of workloads and link loss rates (Raman & McCanne, 1999).

Lui, Misra, & Rubenstein (2004) developed three analytical models to evaluate the robustness of soft/hard state protocols under three network conditions: denial of service attacks, correlated lossy feedback channel, and implosion under multicast services. Simulation results demonstrated that hard state protocols can be optimized to outperform their soft state counterparts under the stable network conditions, but their performance degrades at a much faster rate than that of soft state protocols. Soft state protocols are much more resilient to varying network conditions due to unanticipated fluctuations of call demands in the real signaling network (Lui, Misra, & Rubenstein, 2004).

Hong, Huang, & Yan (2010a) created a Markov-Modulated Poisson Process (MMPP) model to analyze the queuing mechanism of SIP server under two typical service states. The MMPP model can be used to predict the probability of SIP retransmissions, because the theoretical retransmission probability calculated by MMPP model is located within the confidence interval of the real retransmission probability obtained from numerous simulation replications (Hong, Huang, & Yan, 2010a). High retransmission probability caused by short demand surge or reduced server processing capacity during maintenance period may overload and crash a SIP server.

All the above analytical models are useful for studying the statistical performance and predicting the overload probability of a SIP network.

## 3.2. Fluid Model For Overloaded SIP System

As the retransmission mechanism would amplify the SIP overload and make an overloaded SIP server unstable, the analytical models created for a stable SIP system (e.g., Ji, Ge, Kurose, & Towsley, 2007; Lui, Misra, & Rubenstein, 2004; May, Bolot, Jean-Marie, & Diot, 1999; Raman & McCanne, 1999; Zaim et al., 2003) cannot effectively analyze and evaluate the dynamic performance of an overloaded server.

Hong, Huang, & Yan (2010b) developed a fluid model to capture the dynamic behaviour of SIP retransmission mechanism of a single server with infinite buffer. A related study of a tandem server gives the guidance on how to extend the innovative approach to model an arbitrary SIP network (Hong, Huang, & Yan, 2010b). The fluid model can help researchers speed up the performance evaluation using the fluid-based simulation, when extremely high message arrival rate and service capacity of a SIP network are well beyond the computational capabilities of current event-driven simulators (Liu et al., 2003). Numerous simulation results demonstrate that (1) SIP server behaviour is sensitive to the parameters of signaling demands and initial conditions, a characteristic of chaotic systems; (2) overload at a downstream server can propagate to its upstream servers and therefore cause widespread collapse across a SIP network (Hong, Huang, & Yan, 2010b).

On the other hand, the finite buffer would drop messages when a transient overload cause buffer overflow. Retransmission for message loss recovery is non-redundant, while retransmission trigger by long overload delay is redundant. Another novel strategy was introduced to classify different types of retransmission messages so that a fluid model was created for an overloaded SIP tandem server with finite buffer (Hong, Huang, & Yan, 2011a). The impact of finite buffer size on SIP retransmission mechanism was studied in case of the overload. A small buffer size can be a simple overload control mechanism with a cost of arbitrarily high call blocking rate (Hong, Huang, & Yan, 2011a).

## 4. SIP OVERLOAD CONTROL

Recent collapses of SIP servers due to emergency induced call volume or "American Idol" flash crowd in real carrier networks have attracted great research attention and motivated different types of strategies to address SIP server overload problem (Hilt & Widjaja, 2008; Noel & Johnson, 2007; Shen & Schulzrinne, 2010). In order to help researchers gain a comprehensive picture of current proposed SIP overload control solutions, we identify two broad classes of SIP overload control mechanisms: load balancing approach and load reducing approach. Figure 3 depicts the classification for the existing SIP overload control schemes. Load balancing approach aims to avoid the overload by distributing the traffic load equally among the local SIP servers (e.g., Jiang et al., 2009; Warabino, Kishi, & Yokota, 2009), while load reducing approach tries to prevent the overload collapse by reducing the traffic load in the whole SIP network (e.g., Hilt & Widjaja, 2008; Hong, Huang, & Yan, 2010c; Noel & Johnson, 2007; Ohta, 2006a; Shen, Schulzrinne, & Nahum, 2008).

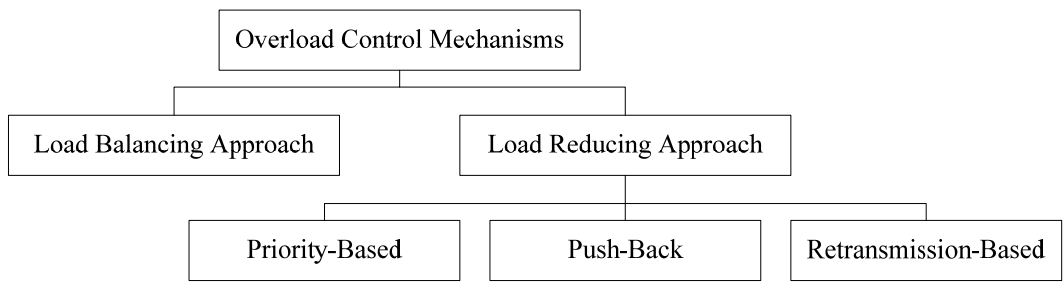

*Figure 3. The classification for the existing SIP overload control schemes*

## 4.1. Load Balancing

The load balancing strategy distributes the newly incoming traffic to each local server based on its available processing capacity, thus reducing the probability that overload happens at a specific server. As a default part of numerous operating systems, SIP Express Router (SER) provides a load balancing module to mitigate the overload caused by large subscriber populations or abnormal operational conditions (IP Telecommunications Portal, 2011). Dacosta, Balasubramaniyan, Ahamad, & Traynor (2009) analyzed the relationship between the latency of message delivery and call throughput in a network configuration where SIP proxies are distributed but the database servers are centralized. Request batching was combined with parallel execution to balance the demands for bandwidth and the call failure rates of a SIP proxy, thus improving call throughput and reduce call failure rate significantly (Dacosta, Balasubramaniyan, Ahamad, & Traynor, 2009).

Jiang et al. (2009) proposed three novel algorithms for balancing load across cluster-based SIP servers. Each algorithm estimates the server load dynamically and performs session-aware request assignment. Call-Join-Shortest-Queue algorithm tracks the number of SIP calls allocated to each back-end server and routes a new incoming call to the server with the least number of active calls; Transaction-Join-Shortest-Queue algorithm routes a new incoming call to the server with fewest active transactions, rather than with the fewest calls; Transaction-Least-Work-Left algorithm routes a new incoming call to the server with the least load, given that different type of transaction consume different cost and the work load is determined by the aggregated estimated cost of the total transactions. All three algorithms exhibit short response time by distributing requests across the cluster more fairly, thus minimizing occupancy and the corresponding waiting time for a particular request behind others for service (Jiang et al., 2009).

Peer-to-Peer network technology was integrated with SIP to balance the traffic load. Warabino, Kishi, & Yokota (2009) proposed session control architecture "Minimum Core", where the core network cooperates with overlay networks to keep the call-setup time as short as possible, while minimizing processing and traffic load on the core network. In the "Minimum Core", small base stations realize autonomous SIP session control using peer-to-peer technology (Warabino, Kishi, & Yokota, 2009). Huang (2009) applied peer-to-peer technology for locating and discovering interested SIP peers. The peers are virtually grouped by classifying their interested topics into n-tuple overlay virtual hierarchical tree in the overlay network. Caching the addresses of peers can prevent the overload of SIP traffic in tree structure (Huang, 2009).

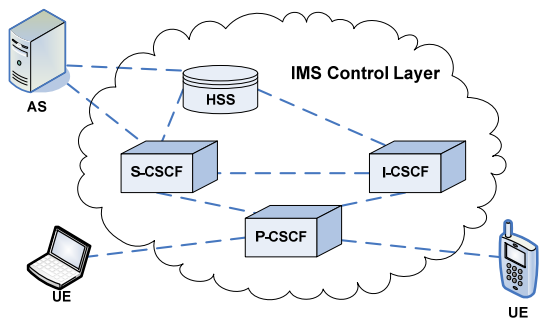

*Figure 4. Simplified IMS control layer overview*

To achieve fixed-mobile convergence (FMC) and ease the integration with the Internet, IMS aids the access of multimedia and voice applications from wireless and wired terminals. Figure 4 illustrates a simplified architecture of the IMS control layer. The main elements of the layer are: the Home Subscriber Servers (HSS) and different types of Call/Session Control Function (CSCF) servers with different purposes. The control layer connects to the Application Server (AS) in the application layer and to the User Equipment (UE) through transport layer (3GPP WG, 2011). A Proxy-CSCF (P-CSCF) acts as the contact point for UE to access the core network service; an Interrogating-CSCF (I-CSCF) provides routing for the signaling messages; and a Serving-CSCF (S-CSCF) takes full responsibility for UE registration, session control, and service routing with AS. An IMS core network provides services for individual end users through a complete set of signaling procedures. The main procedures include UE registration and de-registration, P-CSCF discovery, S-CSCF assignment, session establishment and termination, QoS negotiation and home network directing (3GPP WG, 2006). Due to the stochastic nature of signaling traffic, demand burst can potentially overload certain servers and degrade server performance significantly, which may result in the unrecoverable server collapse (Planat & Kara, 2006). When signaling traffic surge overloads an IMS server which can be S-CSCF or P-CSCF or I-CSCF server, load balancing can mitigate the overload by transferring part of the load to other servers within the same domain.

S-CSCF load balancing – As each S-CSCF server virtually handles all major tasks and processes the majority of signaling traffic, it is vulnerable to overload (3GPP WG, 2011). To subscribe IMS service, each UE initials the registration to associate with a specific S-CSCF. For administrative purpose, each UE is required to perform re-registration periodically. Experimental data analysis by Xu, Huang, Yan, & Drwiega (2009) demonstrates that the increased percentage in registration/de-registration traffic can only slightly affect the overall load distribution over different IMS elements, thus the changes in registration and de-registration rate may affect the consequent total network traffic load, but have limited impact in the general load distribution. In order to avoid further overload at potentially over-utilized S-CSCF, Xu, Huang, Yan, & Drwiega (2009) proposed de-registration based load balancing scheme to re-directs consequent SIP traffic from the over-utilized S-CSCF to the other under-utilized ones. An AS is deployed in the application layer to communicate with S-CSCFs from the core network, and acts as the load-balancing decision maker. By periodically collecting status information (e.g., server utilization) from each S-CSCF, the AS evaluates the S-CSCF load condition and initiates de-registration requests to the overloaded S-CSCF (Xu, Huang, Yan, & Drwiega, 2009).

P-CSCF load balancing – Each P-CSCF routes the SIP requests between the UE and the other S-CSCF/I-CSCF, and establishes a security association with the UE. Once a UE is registered with a specific S-CSCF, the P-CSCF memorizes the SCSCF for the UE (3GPP WG, 2011). Maintaining the periodic registration for a large number of UEs would consume a large amount of computing and memory resources of SIP nodes (Kitatsuji, Noishiki, Itou, & Yokota, 2010). A UE registration scheme proposed by Kitatsuji, Noishiki, Itou, & Yokota (2010) not only balances the workload over multiple P-CSCF nodes, .but also reduces the required P-CSCF nodes up to 40% from the standard session initialization procedure of IMS.

I-CSCF load balancing – Each I-CSCF stores a routing table and decides the message routing to S-CSCFs. In order to serve a large number of requests as efficient as possible, the I-CSCF has to route SIP messages with minimal state information. A SIP message overload transfer scheme proposed by Geng, Wang, Zhao, & Wang (2006) not only provides efficient SIP message routing in high volume I-CSCF servers to balance the load to different S-CSCF servers, but also leverages redundant servers to reduce the message disruption in cases of server failures.

However, load balancing tries to avoid SIP network failures by reducing the utilization of those servers that may become overloaded. When the total message arrival rate exceeds the aggregated processing capacities of all local servers, load balancing schemes cannot prevent the overload collapse.

## 4.2. Priority-Based Overload Control

When all the local servers are experiencing heavy load, priority-based control approaches aim at mitigating the overload by rejecting the calls with low priority.

A priority enqueuing scheme provides differentiate service for different types of SIP messages in every SIP proxy server, where "INVITE" messages are placed into low priority queue and other types of SIP messages are placed into the high priority (Ohta, 2006a). Each proxy server checks both queues alternately. Only when the high priority queue is empty, the server processes "Invite" messages in the low priority queue. Once the proxy server is overloaded, every "INVITE" message would be hardly forwarded to its destination, thus reducing the traffic load by forbidding the successive non-INVITE transactions (Ohta, 2006a). Instead of holding "INVITE" messages in the low priority queue, two thresholds are created for the low priority queue. When the overload drifts the low priority queue over the lower threshold, the proxy server starts to reject the calls with a certain probability (Garroppo, Giordano, Spagna, & Niccolini, 2009).

Amooee & Falahati (2009) leveraged priority queue to overcome the overload problem of an IMS system by blocking non-priority calls. Similar to the priority scheme, Dacosta & Traynor (2010) developed a novel authentication protocol to reduce the load on the centralized authentication database dramatically, while improving the overall security of a carrier-scale VoIP network.

## 4.3. Push-Back Overload Control

Since the CPU cost of rejecting a session is usually comparable to the CPU cost of serving a session (Shen, Schulzrinne, & Nahum, 2008), cancelling "INVITE" transaction using priority queuing scheme is not very cost effective. Therefore, numerous push-back solutions have been proposed to reduce the traffic loads of an overloaded receiving server by advertising its upstream sending servers to decrease their sending rates.

Local overload control mechanism suggests an overloaded SIP server to reject SIP requests locally by sending 503 responses to its upstream servers without Retry-After header (Hilt & Widjaja, 2008). The 503 response code with no Retry-After header would suppress retransmissions of the rejected requests. In bang-bang control algorithm, each server works at underload and overload states alternately: underload state turns to overload state when message queue size exceeds a high threshold; overload state turns to underload state when message queue size falls below a low threshold. All messages are accepted in underload state, while all new arrival INVITE requests are rejected in overload state (Hilt & Widjaja, 2008). In occupancy algorithm, every new arrival INVITE request is rejected with a probability p, where the probability is regulated dynamically to clamp the processor occupancy (or utilization) below a given target value (Hilt & Widjaja, 2008). Rejecting a request consumes less processing resources than fully processing it, however, an overloaded SIP server still needs to utilize most of its processing capacity to reject requests if the newly offered workload is very high. In distributed overload control, an overloaded downstream server informs its overload information explicitly to its upstream sending servers, so that each upstream server can reduce the message sending rate to cancel the overload at the downstream server (Hilt & Widjaja, 2008).

A new retry-after control scheme determines the retry-after timer based on the overloaded proxy load so that the overloaded proxy can drain its input queue to a low level more quickly and significant goodput improvement can be achieved (Noel & Johnson, 2009). For processor occupancy control scheme, an overloaded downstream server calculates a target call rate based on its processor occupancy, and then broadcasts the rate to all its upstream servers (Noel & Johnson, 2009). Similarly, queue delay control scheme performs overload control by calculating a target call rate based on the queuing delay of the overloaded server (Noel & Johnson, 2009). Window-based control scheme allows the edge proxy to forward a new call to a core proxy if and only if the number of outstanding call requests for the core Proxy is strictly less than the corresponding window size. Otherwise the edge proxy rejects the respective call of UAC (Noel & Johnson, 2009).

Three window-based feedback control algorithms proposed by Shen, Schulzrinne, & Nahum (2008) make each downstream receiving server dynamically calculate a window size for its upstream servers based on the overload status. Rate control algorithms implemented by Shen, Schulzrinne, & Nahum (2008) attempt to clamp the message queuing delay or the processor occupancy below a target value. In addition, many other push-back solutions have been proposed to improve the goodput of an overloaded server (e.g., Abdelal & Matragi, 2010; Homayouni, Jahanbakhsh, Azhari, & Akbari, 2010; Montagna & Pignolo, 2010; Ohta, 2006b; Shen & Schulzrinne, 2010; Sun, Tian, Hu, & Yang, 2009; Wang, 2010; Yang, Huang, & Gou, 2009).

The main idea of the push-back control solutions is to cancel the overload of a server by reducing the sending rate of its upstream servers. This would increase the queuing delays of newly arrival original messages at the upstream servers, which in turn cause overload at the upstream servers. Overload may thus propagate server-by-server to sources. Unlike a source in TCP typically generates large amount of data, a UA in SIP only generates very few signalling messages. This leads to rejections of a large number of calls which means revenue loss for carriers. However, it may be unnecessary to reject calls when temporary overload only lasts a short period of time.

## 4.4. Retransmission Rate-based Overload Control

When retransmissions are caused by the overload rather than the message loss, they will bring extra overhead instead of reliability to the network and exacerbate the overload (Sun, Tian, Hu, & Yang, 2009). Through the analysis on the queuing dynamics of two neighboring servers, Hong, Huang, & Yan (2010c) discovered the root cause of the overload propagation which had been verified by numerous experiments (e.g., Hilt & Widjaja, 2008; Noel & Johnson, 2007). Figure 5 depicts the queuing dynamics of two neighboring servers where the initial overload occurs at the downstream Server 2 due to server slowdown (Hong, Huang, & Yan, 2010c). There are two queues at each server: one to store the messages and the other to store the retransmission timers (Hilt & Widjaja, 2008; Shen, Schulzrinne, & Nahum, 2008). The queuing dynamics for the message queue of the downstream Server 2 can be obtained as

$$\dot{q}_2(t) = \lambda_2(t) + r_2(t) + \upsilon_2(t) - \mu_2(t), \tag{1}$$

where $q_2(t)$ is the queue size and $q_2(t) \geq 0$; $\lambda_2(t)$ is original message rate; $r_2(t)$ is retransmission message rate; $\upsilon_2(t)$ is response message rate; $\mu_2(t)$ is the message service rate.

Like Eq. (1), we can obtain the queuing dynamics for the message queue of the Server 1 as

$$\dot{q}_1(t) = \lambda_1(t) + r_1(t) + r'_2(t) + \upsilon_1(t) - \mu_1(t), \tag{2}$$

where $q_1(t)$ is the queue size and $q_1(t) \geq 0$; $\lambda_1(t)$ is original message rate; $r_1(t)$ is retransmission message rate corresponding to $\lambda_1(t)$; $r'_2(t)$ is retransmission message rate generated by Server 1 for $\lambda_2(t)$; $\upsilon_1(t)$ is response message rate corresponding to $\lambda_1(t)$, and the response messages will remove the corresponding retransmission timers from timer queue $q_{r1}$; $\mu_1(t)$ is the message service rate. When Server 2 performs its routine maintenance and reduces its service capacity for signaling messages, the original message rate $\lambda_2(t)$ is larger than the service rate $\mu_2(t)$, the queue size $q_2(t)$ tends to increase according to Eq. (1) (i.e., $\dot{q}_2(t) > 0$). After a short period, the queuing delay of Server 2 is long enough to trigger the retransmissions $r'_2(t)$ which enter the queue of Server 1. If the total new message arrival rate of $\lambda_1(t)$, $\upsilon_1(t)$ and $r'_2(t)$ is larger than the service rate $\mu_1(t)$, the queue size $q_1$ would increase (i.e., $\dot{q}_1(t) > 0$, as indicated by Eq. (2)) and may trigger the retransmissions $r_1(t)$ to bring the overload to Server 1. This indicates the overload propagation from the downstream Server 2 to its upstream Server 1. After queuing and processing delay at Server 1, the retransmitted messages $r'_2(t)$ enter Server 2 as $r_2(t)$ to increase the queue size $q_2(t)$ more quickly (as described by Eq. (1)), thus making the overload at Server 2 worse.

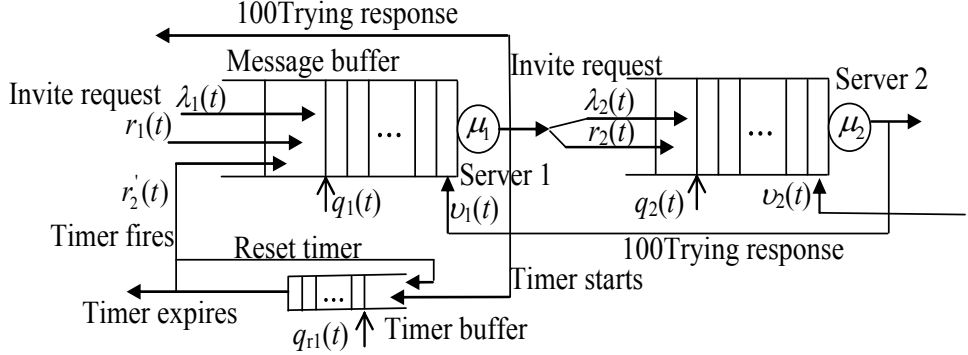

*Figure 5. Queuing dynamics of an overloaded server and its upstream server*

Therefore, while other existing load reducing approaches aim at reducing the original message sending rate of the upstream servers, reducing the retransmission rate can mitigate the overload while maintaining the original message sending rate. The advantages of keeping the original sending rate are less blocked calls and more revenue for the carriers. The key to the retransmission-based solution is to detect overload reliably so that overload can be differentiated from the occasional message loss.

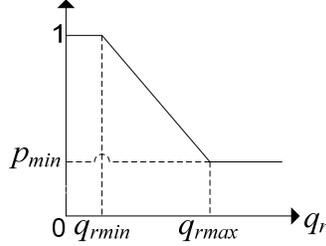

*Figure 6. Retransmission probability vs. Retransmission timer queue size in case of SIP overload*

A solution for the differentiation is as follows. A copy of each original message is placed into a retransmission timer queue at an upstream sending server after it is transmitted to a downstream receiving server. A corresponding response message disables the retransmission timer by removing the original message from the retransmission timer queue. Otherwise, a retransmission would be triggered to recover the potential message loss. When the overload at the downstream receiving server delays the processing of the original message thus the corresponding response, the retransmission timer queue builds up and unnecessary retransmissions are stimulated. Therefore, the queue size of retransmission timer queue was used to detect the overload and a heuristic Retransmission Timer Queue Control (RTQC) algorithm was proposed by Hong, Huang, & Yan (2011b) to mitigate the overload by controlling retransmission rate. When the overload is anticipated at a downstream server, its upstream servers retransmit the messages with a probability based on the instantaneous retransmission timer queue size. Figure 6 depicts the relationship between retransmission probability and retransmission timer queue size. Maximum queue threshold $q_{rmax}$ and minimum queue threshold $q_{rmin}$ are chosen to calculate retransmission probability $p$. Both queue thresholds are adaptively tuned by the average original message departure rate of the upstream server (Hong, Huang, & Yan, 2011b). The packet loss over the Internet causes the message loss directly. Global packet-loss index provided by The Internet Traffic Report (ITR, 2010) indicates that current global packet loss statistic averaged $\zeta$=8% packet loss. A minimum retransmission probability $p_{min}$ is maintained to achieve a low call blocking probability in case of the overload (Hong, Huang, & Yan, 2011b). The main advantage of the heuristic control algorithm is its simple implementation structure. However, when sporadic bandwidth congestion at the TCP layer causes arbitrarily high message loss thus creating a long retransmission timer queue at the SIP layer, the minimum retransmission probability would lead to sluggish message loss recovery.

Only a retransmitted message for message loss recovery is a non-redundant request message as well as an original message, while a retransmission caused by the overload delay is redundant. Correspondingly, a response message corresponding to a redundant retransmitted message is redundant. Thus the redundant retransmission ratio was adopted as the overload indicator (Hong, Huang, & Yan, 2010c). Using a control-theoretic approach, the interaction of an overloaded downstream server with its upstream server was modelled as a feedback control system in Hong, Huang, & Yan (2010c). They proposed the Redundant Retransmission Ratio Control (RRRC) algorithm to keep the redundant retransmission ratio to an acceptable level by controlling the retransmission message rate of its upstream servers, thus mitigating the overload at the downstream server. Since redundant retransmission messages can only be detected after its corresponding response messages are received, such delay might lead to sluggish reaction and potential throughput loss.

As the queuing delay has been well accepted as a more reliable indicator of overload, some existing push-back solutions (e.g., Noel & Johnson, 2007; Shen, Schulzrinne, & Nahum, 2008) attempt to clamp the bottleneck queuing delay below a predefined target queuing delay by reducing the original message rate. The impact of the retransmission rate on the queuing delay of an overloaded server was studied. For the round trip delay between an overloaded server and its upstream server, the queuing delay is dominant, while transmission and propagation delay are negligible (Shen, Schulzrinne, & Nahum, 2008). The upstream server can estimate the round trip delay as the queuing delay of its overloaded downstream server. Hong, Huang, & Yan (2011c) developed the Round Trip Delay Control (RTDC) algorithm to clamp the queuing delay of the overloaded downstream server below a desirable target delay.

Without the modification of the SIP header, the three retransmission-based control solutions (i.e., RTQC, RRRC and RTDC) locate the control algorithm in each upstream server and detect the overload in the downstream server implicitly. In case of short-term overload, the retransmission-based control mechanism can mitigate the overload effectively without rejecting calls or reducing network utilization, thus avoiding the disadvantages of other types of overload control solutions.

## 4.5. Standards on SIP Overload Control

Gurbani, Hilt, & Schulzrinne (2011) have proposed The Internet Engineering Task Force (IETF) Request for Comments (RFC) draft recently to provide the design guideline for SIP overload control mechanism. The Internet-draft defines new parameters for the SIP via header for overload control. These parameters are used to convey overload control information between SIP entities (Gurbani, Hilt, & Schulzrinne, 2011). Different types of the overload control approaches summarized above can be the candidates of future SIP overload control mechanism for potential large scale deployment in the carrier networks.

## 5. CONCLUSIONS AND RESEARCH OPPORTUNITIES

This chapter has briefly reviewed the signaling mechanism of SIP, the dominant signaling protocol in the Internet. The retransmission mechanism for maintaining SIP reliability and the built-in overload control mechanism with limited efficiency have been introduced. Then the chapter has reviewed two main modeling approaches to help researchers understand the root cause of the SIP overload collapse through analyzing theoretically the statistical characteristics and dynamic behaviour of a SIP system. Finally current proposed SIP overload control solutions have been classified into four different types and discussed in details.

Load balancing mechanism and SIP built-in overload control mechanism have been deployed in the carrier networks. Other three types of load reducing approaches are still in the stage of research proposals. Unlike load balancing, priority-based and retransmission-based control approaches, most existing push-back control solutions recommended by SIP overload control RFC (Gurbani, Hilt, & Schulzrinne, 2011) require the modification of the SIP message header in order to explicitly exchange the overload information between neighbouring servers. However, changing the SIP message header requires a time-consuming standardization process and the cooperation among different carriers in different countries. Considering a large scale deployment of SIP servers in the current Internet, new overload control schemes must be incrementally deployable, and slight increases in complexity or cost may be strongly resisted by the carriers. The solutions without modifying protocol header may lure more interests from the carriers due to the ease of quick implementation.

Resource over-provisioning can reduce the overload probability significantly, but such passive action would cause low average capacity utilization and increase the capital costs. Rejecting calls can mitigate the overload quickly, but it would reduce the revenue and decrease user satisfaction index. It may be unnecessary to reject calls upon a short-term overload.

In the future work, carriers and researchers can cooperate to evaluate the efficiency of different SIP overload control approaches by performing field test in the carrier networks. To achieve a good trade-off between the revenue and the response time to cancel the overload in different overload scenarios (e.g., short-term overload and long-term overload), more research is needed to determine how to choose and combine different types of overload control approaches to avoid the potential SIP overload collapse.

**Citation of this book chapter**:

Y. Hong, C. Huang, and J. Yan, "A Comparative Study of SIP Overload Control Algorithms," *Network and Traffic Engineering in Emerging Distributed Computing Applications*, Edited by J. Abawajy, M. Pathan, M. Rahman, A.K. Pathan, and M.M. Deris, IGI Global, 2012, pp. 1-20.

**Note**:

(1) Explicit SIP overload control algorithm requires the modification in the SIP header and the cooperation among different carriers in different countries.
(2) Implicit SIP overload control algorithm does NOT require the modification in the SIP header and the cooperation among different carriers in different countries. Any carrier can freely implement implicit SIP overload control algorithm in its SIP servers to avoid potential widespread server crash.
(3) Due to the copyright agreement, this is the original draft of the book chapter. The final draft of the book chapter can be found in the following IGI Global link. The difference between both the original draft and the final draft is that the final draft includes the minor revision suggested by the peer reviewers.
http://www.igi-global.com/chapter/comparative-study-sip-overload-control/67496
(4) OPNET simulation codes for 3 implicit SIP overload control algorithms (RRRC, RTDC, and RTQC) published by *Telecommunication Systems* and *Proceedings of IEEE Globecom 2010/ICC 2011* are available for non-commercial research use upon request. Presentation slides for RRRC and RTDC algorithms can be downloaded from the following RG links.
https://www.researchgate.net/publication/258555827_Mitigating_SIP_Overload_Using_a_Control-Theoretic_Approach
https://www.researchgate.net/publication/257945199_Round-Trip_Delay_Control_(RTDC)_For_Mitigating_SIP_Overload_(IEEE_ICC_2011_Slides)
(5) RRRC algorithm has been quickly adopted by The Central Weather Bureau of Taiwan for their early earthquake warning system: "An Efficient Earthquake Early Warning Message Delivery Algorithm Using an in Time Control-Theoretic Approach," Ubiquitous Intelligence and Computing, Lecture Notes in Computer Science, 6905, Springer, Berlin, Heidelberg, 2011, pp. 161-173.
http://www.springerlink.com/content/b6252x2k613rv211/?MUD=MP
http://www.ipv6.org.tw/docu/elearning8_2011/1010004798p_3-7.pdf
Short review and comments on RRRC algorithm: "Local SIP Overload Control," *Proceedings of WWIC*, June 2013.
http://link.springer.com/chapter/10.1007%2F978-3-642-38401-1_16#
http://c3lab.poliba.it/images/2/2a/SipOverload_WWIC13.pdf
(6) RTDC algorithm has been recommended as White Paper by TechRepublic (an online trade publication and social community for IT professionals, part of the CBS Interactive).
http://www.techrepublic.com/whitepapers/design-of-a-pi-rate-controller-for-mitigating-sip-overload/25142469
(7) Journal paper not only conducts more theoretical analysis of Round trip delay control (RTDC) and Redundant retransmission ratio control (RRRC), but also discusses how to apply RTDC algorithm to mitigate SIP overload for both SIP over UDP and SIP over TCP (with TLS): Y. Hong, C. Huang, and J. Yan, "Applying control theoretic approach to mitigate SIP overload," *Telecommunication Systems*, **54**(4), 2013, pp. 387-404.
https://www.researchgate.net/publication/257667871_Applying_control_theoretic_approach_to_mitigate_SIP_overload
http://link.springer.com/article/10.1007/s11235-013-9744-8
(8) Discussion on control system design can be found in the answers to the ResearchGate question "What are trends in control theory and its applications in physical systems (from a research point of view)?", as shown in the following RG link.
https://www.researchgate.net/post/What_are_trends_in_control_theory_and_its_applications_in_physical_systems_from_a_research_point_of_view2